\documentclass[12pt,tightenlines,superscriptaddress,amsmath,amssymb,prl]{revtex4-1}
\usepackage{pdfpages} 
\usepackage{mathptmx}
\usepackage[utf8]{inputenc}
\usepackage{geometry}
\usepackage{wrapfig}
\usepackage{amsmath}
\usepackage{amssymb}
\usepackage{stmaryrd}
\usepackage{graphicx}
\usepackage{etoolbox} 

\newcommand*\LyXThinSpace{\,\hspace{0pt}}

\usepackage{times}
\usepackage{latexsym}\usepackage{epsfig}
\usepackage{lineno}
\usepackage{fancyhdr}

\usepackage{dcolumn}
\usepackage{bm}
\usepackage[compact]{titlesec}
\titlespacing{\section}{0pt}{2ex}{1ex}
    \titlespacing{\subsection}{0pt}{1ex}{0ex}
    \titlespacing{\subsubsection}{0pt}{0.5ex}{0ex}
\usepackage{color}

\newcommand{\msun}{{\rm M}_\odot}

\newcommand{\lsim}{\mathrel{\rlap{\lower4pt\hbox{\hskip1pt$\sim$}}
        \raise1pt\hbox{$<$}}}
\newcommand{\gsim}{\mathrel{\rlap{\lower4pt\hbox{\hskip1pt$\sim$}}
        \raise1pt\hbox{$>$}}}

\makeatletter
\patchcmd{\@outputpage@head}{\@ifx{\LS@rot\@undefined}{}{\LS@rot}}{}{}{}
\makeatother

\begin{document}

\includepdf{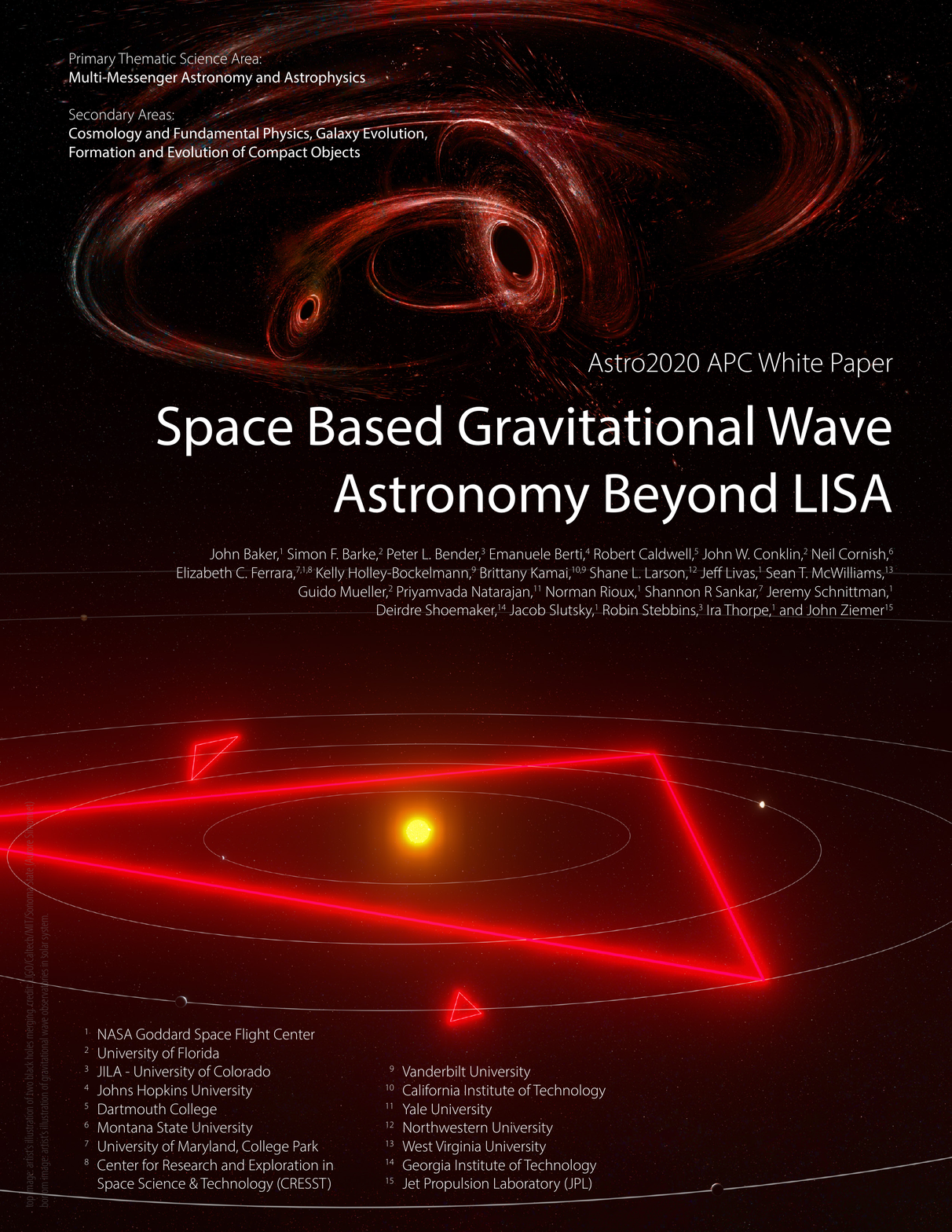}
\nopagebreak
\pagenumbering{gobble}
\setcounter{page}{1}
\pagenumbering{arabic}

\section{Executive Summary}

The Laser Interferometer Space Antenna (LISA) will open three decades
of gravitational wave (GW) spectrum between 0.1 and 100\,mHz, the
mHz band \citep{LISAProposal2017}. This band is expected to be the
richest part of the GW spectrum, in types of sources, numbers of sources,
signal-to-noise ratios and discovery potential. When LISA opens the
low-frequency window of the gravitational wave spectrum, around 2034,
the surge of gravitational-wave astronomy will strongly compel a subsequent
mission to further explore the frequency bands of the GW spectrum
that can only be accessed from space. The 2020's is the time to start
developing technology and studying mission concepts for a large-scale
mission to be launched in the 2040's. The mission concept would then
be proposed to Astro2030.

Only space-based missions can access the GW spectrum between $10^{-8}$
and 1\,Hz because of the Earth's seismic noise. This white paper
surveys the science in this band and mission concepts that could accomplish
that science. The proposed small scale activity is a technology development
program that would support a range of concepts and a mission concept
study to choose a specific mission concept for Astro2030. In this
white paper, we will refer to a generic GW mission beyond LISA as
\emph{bLISA.}

\section{Advancing mHz Gravitational Wave Astronomy Beyond LISA}

Gravitational Wave astronomy is a new and promising field. LIGO \citep{aLIGO15}
has shown that gravitational-wave observatories can make routine observations
of sources that are invisible to electromagnetic (EM) observations,
and of sources for which complementary EM and GW information is extraordinarily
powerful. In the first ever observation of GWs, GW150914 \citep{GW150914},
LIGO demonstrated the discovery potential of GW observations, by detecting
merging stellar black holes with masses substantially higher than
expected. The subsequent observation of a merging neutron star binary,
GW170817 \citep{GW170817}, that was also widely observed across the
EM spectrum simultaneously showed the power of coordinated observations
and the power of multiple detectors (i.e., LIGO Hanford, LIGO Livingston,
and Virgo). The extraordinary campaign of EM observations that followed
GW170817 provided a tremendous impetus to multi-messenger and time
domain astronomy.

\begin{figure}
\noindent \begin{centering}
\includegraphics[width=0.9\textwidth]{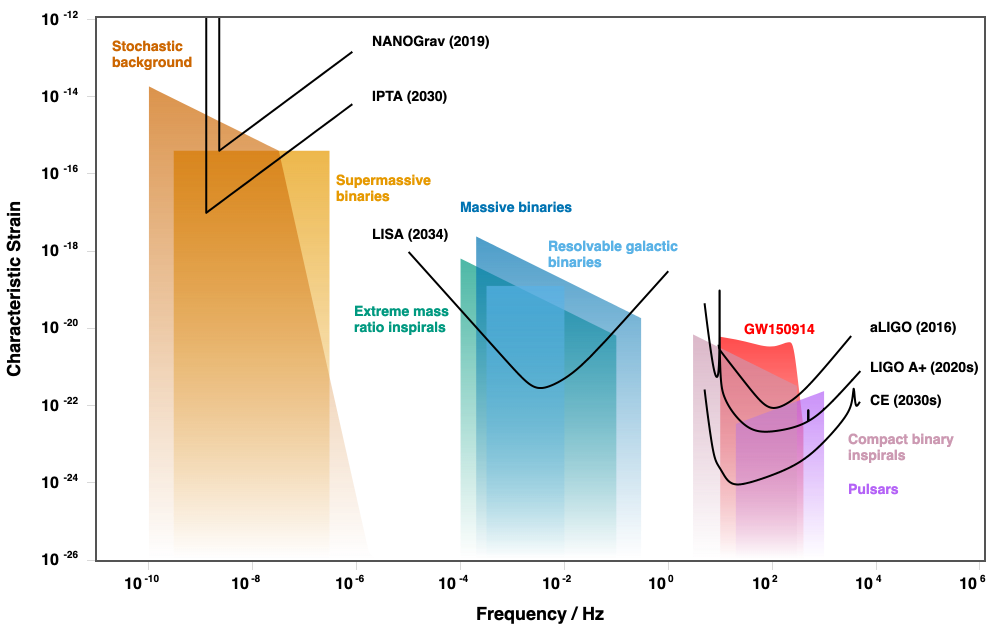} 
\par\end{centering}
\caption{GW Spectrum, showing frequency bands where PTAs, LISA and ground-based
GW detectors operate (generated by http://gwplotter.com/, see \citep{GW_Plotter}
for details).}

\noindent \centering{}\label{fig:GWspectrum} 
\end{figure}

The GW spectrum in Figure~\ref{fig:GWspectrum} shows the detection
strategies for GW astronomy. Roughly speaking, GW sources are inspiraling
binaries (or triples) of compact objects whose GW frequency chirps
up to a final merger frequency, followed by ring-down of the final
object. The gravitational wave frequency roughly scales inversely
with system mass. Pulsar timing arrays (PTAs) detect the largest SMBHs
over a decade of frequency in the nHz region, and ground-based interferometers
detect stellar-mass systems between 40 and 1000\,Hz. LISA spans 0.1
to 100\,mHz. All three strategies are amplitude – not power – detectors,
meaning signal falls off as $1/\text{R}$, rather than $1/\text{R}^{2}$.

Based on the success of LISA Pathfinder \citep{LPF18a,LPF18,LPF18b,LPF19a,LPF19b,LPF19c}
ESA is leading the LISA mission to open up the mHz GW band, with NASA
as a junior partner. This band is expected to have many more detectable
types of sources than any other in the GW spectrum: $10^{4}-10^{7}\msun$
massive black hole binaries (MBHBs), intermediate mass black hole
binaries (IMBHBs), extreme-mass-ratio inspirals (EMRIs) with mass
ratios of $10^{4}$ to $10^{5}$ with a system mass $<10^{7}\msun$,
intermediate-mass-ratio inspirals (IMRIs) with smaller system masses
of $10^{3}-10^{4}\msun$, close compact binaries in the Milky Way,
and the heavy stellar binaries (10's of $\msun$) seen by LIGO and
Virgo. Larger sources will be detectable back into the re-ionization
era. The number of sources is expected to be 10's of thousands. Signal-to-noise
ratios (SNRs) can range into the thousands for the strongest signals.
Further, the detectable mass of sources at cosmological distances
is augmented by their redshift, a substantial boost at, say, $z\sim10$.
LISA's discovery potential includes cosmic strings, cosmological phase
transitions, unexpected bursts and a stronger cosmological GW background
than standard inflationary models predict. LISA's impact on GW astronomy
will parallel LIGO's impact and a space-based GW detector to follow
LISA will be critical to NASA's portfolio.

Below 10\,nHz, pulsar timing arrays (PTAs) observe through long-term,
precisely timed observations of rapidly rotating radio pulsars. It
is widely anticipated that they will measure signals from the largest
MBH binaries $(\sim10^{9}\msun)$ at low redshifts in the 2020's.
PTA sensitivity improves with the discovery and long-term observation
of highly stable pulsars \citep{IPTA16}.

At the high frequency end of the spectrum (40-1000 Hz), ground-based
GW detectors like LIGO and Virgo are currently alternating between
observations that are producing ever more sources, and upgrades leading
to increased sensitivity and duty cycle. Increased sensitivity improves
SNR and enlarges the observable volume as the cube because these are
amplitude detectors. Improved seismic isolation and suspensions are
extending the useful observing band down towards 10\,Hz. During the
current observing run, slated to end in 2020 after a year of operation,
LIGO and Virgo are releasing detection alerts about once per week.
Upgrades are scheduled to continue for several years. Sky localization
will continue to improve as new detectors, KAGRA \citep{Kagra} in
Japan and LIGO-India, come online over the same time scale. So-called
3G detectors \citep{EinsteinTelescope,3G}, with longer armlengths,
improved suspensions and other technology improvements, are in the
early planning stages, with implementation notionally in the 2030s.
These detectors would further improve sensitivity, extend their operating
band down towards 1\,Hz and capitalize on the benefits of a global
network for the best sky localization.

To advance beyond LISA, we survey the science goals and then consider
two illustrative examples of mission concepts. So far in advance of
LISA, it is unwise to focus on a single design. It is more prudent
to examine a range of designs, evaluate the technology needs, some
of which are shared, and down-select to a concept when the technology
challenges are better understood and science priorities are more refined.
The schedule advanced here is for technology development and a mission
concept study in the 2020s leading to a concept selection in time
to propose to Astro2030. Ideally, bLISA would start implementation
when LISA is in its extended mission.

There have been previous efforts to examine possible missions succeeding
LISA. One of the earliest was the Big Bang Observer \citep{BBO2003},
a visionary concept to detect the cosmological GW background. Crowder
and Cornish \citep{CrowderCornish05} compared LISA, BBO, ALIA \citep{ALIABender13}
(to be described below) and stereo versions of LISA and ALIA.

In 2012, a NASA study team produced a wide-ranging study of alternative
designs to LISA \citep{NASASGO12}. While this study did not address
science beyond LISA science, it did comprehensively survey the architectural
choices for space-based GW detectors. The low-frequency Folkner design
described below was derived from this study. The GADFLI \citep{GADFLI}
and GEOGRAWI (now published under the acronym gLISA) \citep{gLISA}
concepts, mentioned below, also originated from this study.

\section{Key Science Goals}

Similar to the EM-spectrum, GW sources are present throughout the
universe and emit over many decades in frequency space. GW emission
is directly linked to the mass of the emitting system; heavier binary
systems merge at lower frequencies while lighter masses pass through
the low frequency band, often with detectable amplitudes, and merge
then at higher frequencies. Closing frequency gaps in the observed
spectrum has two distinct scientific ramifications: (1) mergers can
be measured across (nearly) all mass ranges and out to very large
redshifts; and (2) the inspiral phase can be tracked over many months
and often over many years prior to the merger. Both improvements provide
unique opportunities to test GR. At least as important are improvements
in the 3D localization of these sources to enable coordinated EM observations.

Improving angular resolution is an age-old goal for many areas of
astronomy, and this may be especially important for gravitational
wave astronomy, where the GW signal unequivocally identifies the source
while EM and particle observatories identify the EM/particle signatures
of these sources. LISA localizations may typically be on the order
of a few square degrees (varying significantly for different sources).
With incident wavelengths measured in millions of km, the Rayleigh
criterion $\delta\theta\leq\lambda/D$ indicates the challenge of
directly resolving gravitational wave sources. A large fraction of
GW science, however, relies primarily on astrometric location of point
sources, enhanced beyond the resolving power by the SNR ($\rho$).
So increases in sensitivity are less important for a census of the
black hole population but are crucial to improve the angular resolution
and enable multi-messenger observations.

A key detection milestone will be to achieve angular (and distance)
location with sufficient precision to localize extragalactic events
such as EMRI's and merged massive or super-massive BHB systems to
a single galaxy (or cluster). Approximate localization of these BHB
systems to within arcminutes in advance of merger would enable deep
coincident observations to identify an EM counterpart and thus locate
the galaxy. These multi-messenger observations would then also enable
studies of accretion flows and physics along with details of galaxy
properties, all with precise knowledge of the black hole masses, spins
and recent merger history. Some MBHB galaxy identifications may be
possible with LISA, but future GW missions, in concert with advanced
EM facilities should be able to dramatically expand the rich multi-messenger
data set to enable a robust understanding of the roles black holes
and galaxies play in shaping each other through accretion and mergers.

\subsection{The mHz to Hz frequency band}

The mHz to Hz frequency band is the band in which intermediate mass
black hole binaries, even from lower-mass MBH seeds from Pop-III stars,
would merge. A future observatory with significantly improved sensitivity
compared to LISA provides clear statistics on these still uncertain
objects. Beyond proving their existence and identifying their role
between stellar mass BHs and massive BHs, precise measurements of
the phasing of these mergers would probe for extensions of GR, testing
the presence of additional physical fields with (effective) mass (e.g.
ultralight fields/dark matter candidates).

Another unique source for this frequency range are (typically extra-galactic)
double white dwarf (DWD) systems which could be observed at periods
around their point of initial contact, allowing for example the observation
of a gravitational-wave signal accompanying a type Ia supernova. Based
on the known rate of type Ia supernovae, this also requires significant
improvements in the high frequency sensitivity of LISA. Such a DWD-merger
signal would provide unprecedented early warning for a supernovae.

Observations in this band also provide improved and advanced localization
of the stellar-mass NS and BH chirping binaries with mergers observed
by 3G ground-based GW observatories. Mergers would be precisely forecast
enabling unique options for multi-messenger observations. If intermediate
mass BHs exist in the universe, joint observations of these systems
on the ground and in space will allow us to localize these sources
well before they merge and to break parameter estimation degeneracies.
At sufficiently high sensitivities it could be possible to independently
resolve all such systems to high redshift, providing a complete census
of these systems, while clearly exposing any primordial background
like that predicted by some models of inflation.

LISA will be able to detect extreme mass-ratio inspirals (EMRIs) composed
of a $10\,\msun$ black hole falling into a $3\times10^{5}\msun$
BH to as far as $z=3$, but this horizon decreases rapidly for different
BH mass, or for smaller secondary masses. While the rates of these
remain quite uncertain, we might expect LISA to detect a few per year
based on current models, but increased sensitivities just above a
mHz would enhance the rate roughly as $\rho^{3}$. Plausible sensitivity
improvements would provide a census of the EMRI population to moderate
redshift, providing insights into the dynamics in stellar clusters.
More sensitive observations of EMRIs would enable high precision spacetime
mapping to verify the Kerr nature of astrophysical black holes. For
the strongest EMRI signals we might even gain access to the overtones
in the ring-down radiation. 

\subsection{The sub-$\mu\text{Hz}$ to mHz band}

\begin{wrapfigure}{l}{0.6\textwidth}%
\noindent \centering{} \includegraphics[width=0.6\textwidth]{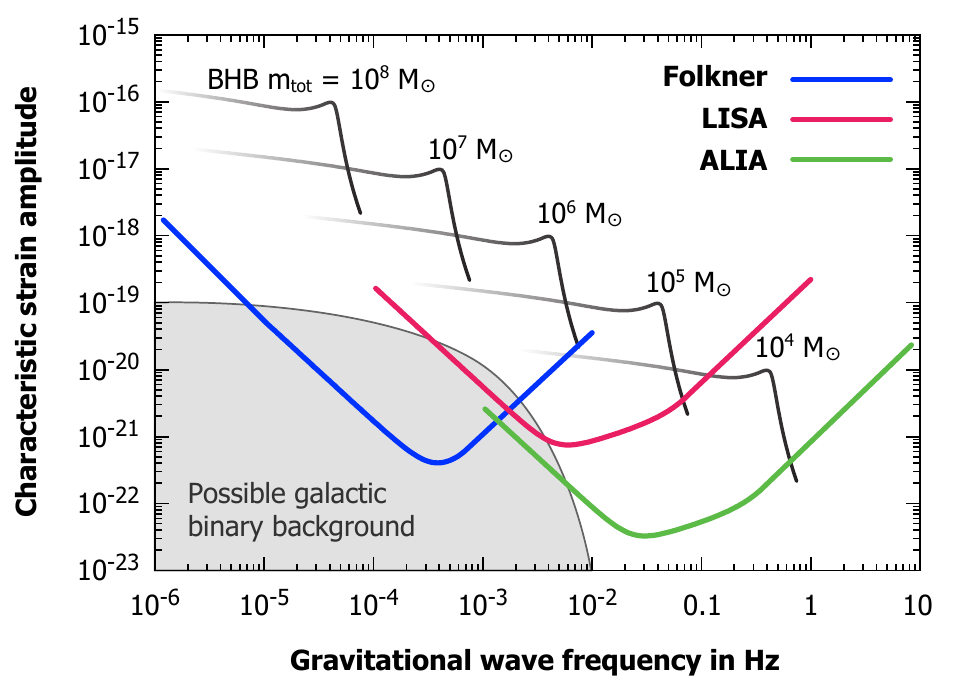}
\caption{Two potential future mission designs compared to the current LISA
design. The low frequency, or Folkner design, assumes that we can
extend LISA's acceleration noise into the low frequency band while
the high frequency design (ALIA) assumes a frequency independent order
of magnitude improvement in the current acceleration noise. The graph
also shows the traces of a few equal-mass black hole mergers as examples
of the science return of each of these missions. The grey shaded area
is a potential representation of the still unknown stochastic gravitational
wave emission from the millions of galactic binaries.}
\label{fig:sensitivity} \end{wrapfigure}%

A similar gain in sensitivity and therefore angular resolution in
the sub-mHz frequency band is frustrated by the stochastic background
of the myriad of GW signals from galactic binary systems. The exact
shape of this spectrum is subject to debate; current and future survey
missions as well as LISA will provide some answers over the next 20
years. However, it is expected that the sensitivity of LISA might
be limited by this background below about one mHz. The grey area in
Figure \ref{fig:sensitivity} shows one of many possibilities for
this background. \emph{One person's noise is another person's signal.}
Any future space mission probing below the LISA band will be able
to measure the spectral and spatial distribution of these binaries
over more than two decades in frequency space and will be able to
identify and isolate probably hundreds of thousands of the stronger
binary systems in our galaxy.

The different models of the galactic binary background still allow
detection of gravitational waves from merging super-massive black
holes in the $10^{7-9}\msun$ mass range out to large redshifts. These
signals will not only map out the Kerr nature of super-massive black
holes with high SNR but also shed light on many of the most fundamental
questions in cosmology and galaxy evolution.

One example is the still unsolved puzzle of the rapid emergence of
the high-z QSOs with big implications not only on the cosmic evolution
of super-massive black holes but also on their impact on early galaxy
formation. Do they regulate the entropy of the intergalactic medium,
hence the fuel of galaxy formation? LISA will probe the $10^{4-6}\msun$
seeds of the early QSOs but will not tell us how and how fast they
reached $10^{8-9}\msun$ (e.g., by Super-Eddington accretion). At
high z it is expected that merger rates of galaxies and subsequent
mergers between their central MBHs is very high. The observation of
the generated GWs is a good probe of the growth function and is probably
the only tool for a reliable census of MBHs during this epoch. In
the mid range of this low frequency detector, we could possibly see
thousands of inspiraling MBHBs with periodicities of days accessible
with future (post LSST) time domain surveys. GWs would provide the
luminosity distance and the EM-observatories the redshift for these
standard sirens. These multi-messenger observations would also provide
powerful insight onto accretion on binaries, the physics of disk-binary
interaction, and many other aspects which need the masses and spins
of the central engines behind these sources. Sensitivity at the high
frequency end of this detector enables earlier detection and localization
of LISA's MBHBs which will greatly expand opportunities for multi-messenger
observations before and during merger. 

\section{Example Mission Concepts}

Looking to the future, we need to explore the science opportunities
beyond LISA, and also the technological hurdles necessary to realize
a beyond-LISA mission. Figure\,\ref{fig:sensitivity} compares the
sensitivity curves of two example mission designs to LISA. These designs
were chosen to illustrate the range of mission concepts, and to show
what is possible if a healthy technology development program for future
gravitational wave missions is established parallel to the LISA project
itself.

\begin{wrapfigure}{r}{0.4\textwidth}%
\noindent \centering{} \includegraphics[width=0.4\textwidth]{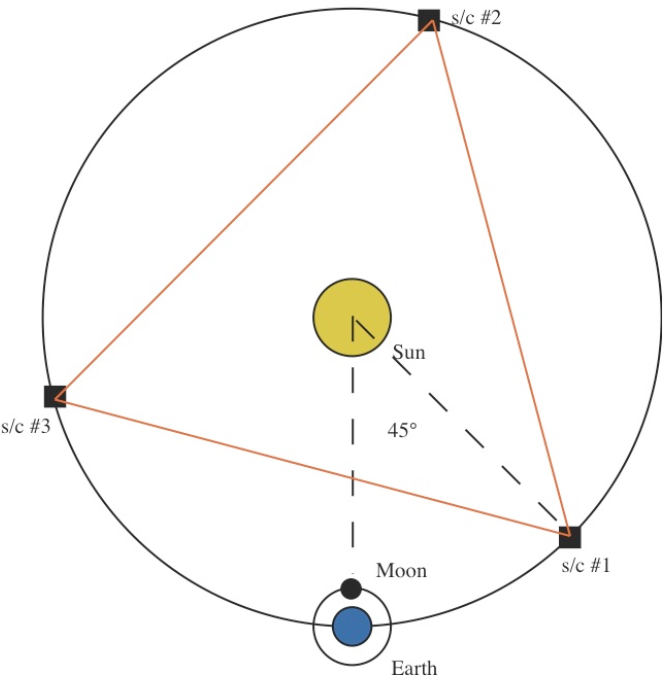}
\caption{The Folkner low-frequency mission concept placed three spacecraft
in this triangular configuration around the Sun. }
\label{fig:orbits} \end{wrapfigure}%

One mission design explores the frequency band below LISA and is based
on orbits in which the three spacecraft form an equilateral triangle
with the Sun in its center. The arms are about 100 times longer than
the LISA arms. This design keeps one spacecraft in the vicinity of
Earth to simplify communication with the constellation. Communication
to the other spacecraft would probably have to be carried by the laser
links. These orbits were originally suggested by William Folkner during
the GW Mission Concept Study \citep{NASASGO12} prior to the LISA
Pathfinder success. He predicted some sensitivity to large massive
black hole merger systems in a mission without inertial sensors where
each spacecraft would act as a test masses. We added a drag free system
back into the spacecraft and assumed that it maintains a frequency
independent acceleration noise identical to the LISA requirement.

The second mission design, the Advanced Laser Interferometer Antenna
(ALIA) \citep{ALIABender13}, assumes a ten times improved acceleration
noise in the high frequency part of the LISA spectrum. This example
mission is five times shorter than LISA but uses significantly more
laser power and larger telescopes to increase the interferometric
sensitivity. The orbits for ALIA are still heliocentric but other
mission designs with shorter arms explored also geocentric orbits
(GADFLI and GEOGRAWI, a.k.a. gLISA).

\subsection{Low-Frequency Mission Example}

The orbits proposed in the Folkner mission would separate the spacecraft
by around 260\,Gm (see Fig.\,\ref{fig:orbits}). One of them has
to be kept at a reasonable distance to Earth to allow for spacecraft
to ground communication which means that lasercom between the spacecraft
has to be part of the mission design.

Folkner also provided an initial study of the orbital dynamics for
the first five years of the mission. He found that differential velocities
will stay below 3m/s, which affects the Doppler shifts and therefore
the frequencies of the beat signals between the laser beams. Furthermore,
changes in the opening angles between the spacecraft stay below $0.02^{\circ}$
and would likely not require in-flight adjustments of the opening
angle between the two outgoing beams on each spacecraft.

For the long arm interferometer, we used parameters which are very
similar to LISA: 30~cm telescopes and a 3\,W laser system. The received
power at the far end would be approximately 150\,fW and the shot
noise limited displacement sensitivity would be a factor 200 below
LISA leading to the same strain sensitivity when expressed as a linear
spectral density; the sensitivity curves in Figure\,\ref{fig:sensitivity}
are scaled by $\sqrt{f}$ to take into account the longer observation
time for low-f signals. As shown in Figure\,\ref{fig:sensitivity},
the galactic binary background will likely limit the sensitivity well
before shot noise. 

\subsection{High-Frequency Mission Example}

We use the ALIA mission design as an example for a high frequency
beyond LISA mission. ALIA's orbits are LISA-like heliocentric orbits
trailing or leading Earth by a few degrees. The distance between the
spacecraft is 500,000\,km; five times shorter than LISA. The resulting
loss in low frequency sensitivity is (over-) compensated by an assumed
ten-fold improvement in acceleration noise. These two parameters can
be fine-tuned once the galactic binary background radiation has been
characterized by LISA. The main improvement comes from a 40,000-fold
increase in the received laser power which ALIA achieves by increasing
the telescope diameter to 1\,m and the laser power to 30\,W. This
leads to a factor 200 improvement in phase/displacement sensitivity
and, due to the factor 5 shorter arms, to a 40-fold improvement in
strain sensitivity. The dynamics of the orbits, relative spacecraft
velocities and angular changes, scale with the arm length (assuming
all other parameters stay the same) and will be reduced by a factor
five compared to LISA.

The spacecraft in the geostationary mission designs were separated
by 73,000\,km. The shorter arm length shifts the sensitivity curve
further to the right. Using ALIA's laser power and telescope diameter
would give the same strain sensitivity at higher frequencies; the
increase in received laser power is compensated by the decrease in
length. However, it requires further improvements in the phase sensing
system. The advantage of geostationary orbits is the reduced launch
costs and simple ground to space communication links. On the other
hand, geostationary constellations likely require station-keeping
to manage the relative velocities (Doppler shifts) and changes in
the opening angles.

\section{Technology}

Space-based GW detectors, based on laser interferometry, generally
have two noise regimes that determine the performance. Residual accelerations
on the inertial reference masses from unwanted disturbances limit
the low frequency performance. Displacement measurement noise from
the interferometry limit the high frequency performance. Together
with the frequency response of a chosen armlength to the GW wavelength,
these two noise types give the bucket-shaped noise curves that characterize
laser-interferometer-based detectors. Any technology improvements
naturally have to address these noise types. Changing a detector's
operating band also requires adjustments to the treatment of these
noise types.

\subsection{Acceleration Noise}

LPF improved on existing drag free systems by several orders of magnitude
and represents the state of the art in force free motion. The LISA
Pathfinder team has done a tremendous job of understanding and characterizing
the limiting noise sources of their gravitational reference sensor
(GRS) which is now baselined for LISA. However, the initial concept
was born in the mid '90s based on a design sensitivity which was defined
prior to understanding the true limitations of such a system. Now
we have a much better understanding of these limitations and, with
that knowledge, should be able to take a fresh look at the design
and find ways to improve it.

\begin{wrapfigure}{r}{0.6\textwidth}%
\noindent \centering{} \includegraphics[width=0.6\textwidth]{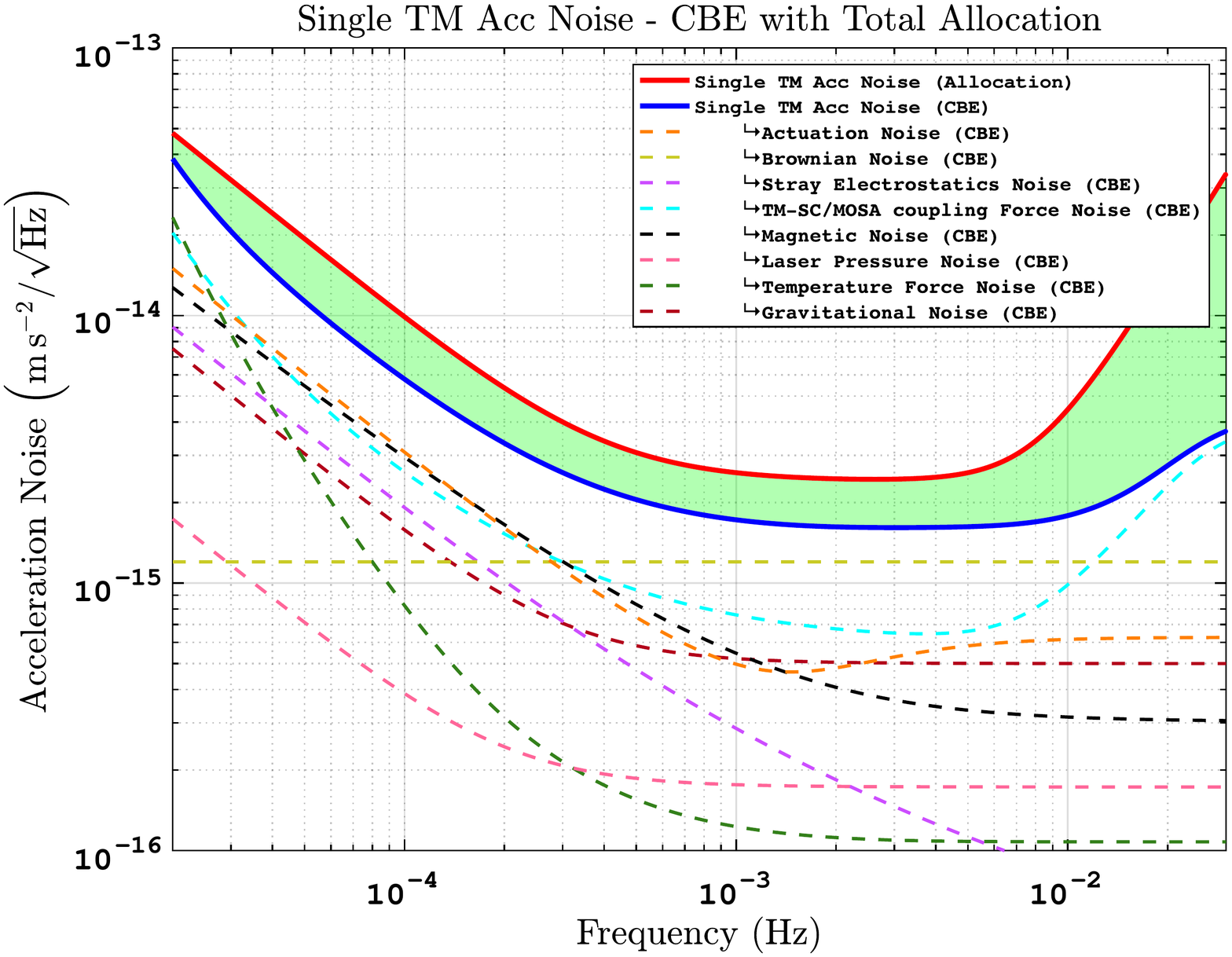}
\caption{The CBE of LISA's acceleration noise is based on LPF data and offers
opportunities for future improvements \citep{LISAPerfModel}.}
\label{fig:acceleration_noise} \end{wrapfigure}%

The LPF team has analyzed hundreds of different potential noise sources.
The leading contributors together with the LISA requirement and current
best estimate are shown in Figure \ref{fig:acceleration_noise} \citep{LISAPerfModel}.
At high frequencies, the noise in the actuation forces between the
spacecraft and the test mass limit the performance. These forces are
typically described as a stiffness along each translational and rotational
degree of freedom. In a perfect drag free system, the spacecraft is
tracking the motion of the test mass and the forces would be constant.
A real drag free system is limited by sensor noise; in LISA we assume
$\sim\text{nm}/\sqrt{\text{Hz}}$ and a few hundred $\text{nrad}/\sqrt{\text{Hz}}$
noise in the capacitive sensors, and by the response time of the $\mu\text{N}$-thrusters.
The later is responsible for the increased noise at higher frequencies.
A related noise source is gravitational noise or changes in the local
gravitational field due to spacecraft motion. At high frequencies,
this is again coupled to the non-suppressed spacecraft motion mostly
along the non-sensitive axis due to cross coupling between the different
degrees of freedom. In LISA, these degrees are only sensed using the
capacitive sensors. One possible improvement in future missions would
be to add a much more sensitive interferometric sensing system to
monitor all degrees of freedom of testmass to spacecraft motion and
not only the one along the optical axis. The design of such a system,
the required sensitivity along the other degrees of freedom, added
complexity and all ramifications with respect to the payload design
need to be explored. 

At lower frequencies, thermo-elastic deformations of the spacecraft
will also contribute to the overall noise budget. This noise falls
off with distance cubed. Potential needed changes include improved
thermal stability which has been identified as one of the driving
forces behind many of the low frequency noise terms. In addition to
passive shielding, LISA will also actively stabilize the temperature
of the outer thermal shields which encapsulate the sensitive parts
of the payload. The requirements on this shielding and the active
temperature stabilization system has to be extended towards much lower
frequencies; again a potentially solvable problem with significant
design implications. A parallel approach is to use only low-CTE materials
(with low water content) in the vicinity of the test mass to minimize
thermo-elastic deformations. Low-CTE materials in this context might
not be restricted to ULE or Zerodur but might also include support
structures build from negative and positive CTE materials of similar
density.

The limiting force in LPF in the most sensitive frequency range is
Brownian noise caused by residual gas molecules bouncing off the test
mass. This frequency independent Brownian motion scales with the area
to mass ratio of the test mass, the residual gas pressure which in
LPF was estimated to be $\sim2\,\mu\text{Pa}$ at the end of life,
and on the gap size between test mass and housing. LISA carries a
requirement of $1\,\mu\text{Pa}$ to provide additional margin, which
is still well above the ultra-high vacuum pressure that can be achieved
in modern laboratories and many orders of magnitude away from the
residual pressure in orbit. The gap size was optimized to reduce this
gas pressure noise while maintaining enough sensitivity in the capacitive
sensors and control authority in the electrostatic actuators. 1/f-noise
in the electrostatic actuators was one of the dominant frequency dependent
noise sources limiting LPF at low frequencies. Alternative mission
concepts might use a single test mass which is gravitationally balanced
within the center of the spacecraft. If feasible, such a design could
significantly reduce the needs for test mass actuation and allow to
increase the gap size by maybe even an order of magnitude. Obviously,
such a design would again benefit from an all interferometric sensing
system but other aspects such as redundancy are a major concern.

Another noise source, named stray electrostatic noise in Figure\,\ref{fig:acceleration_noise},
is related to the test mass charge and electrostatic fields. LPF used
a discontinuous UV discharging system which was turned on regularly
to keep the charges below a few million electrons. LISA is planning
to use a continuous UV discharging system which is expected to reduce
the test mass charge significantly and reduce this noise as well.
However, further improvements in the charge sensing technique might
be needed to reduce the residual charge on the test mass again. The
current charge sensing system relies on the electrostatic actuators
which would limit the proposed increase in gap size to overcome other
noise sources.

One example for a noise that scales with the volume of the test mass
is magnetic force noise. Changing magnetic fields couple to the non-vanishing
magnetic susceptibility of the test mass material and to the magnetic
dipole moment from ferromagnetic inclusions. A specific gold-platinum
alloy was chosen because of its vanishing magnetic susceptibility.
A third and, based on LPF experience, likely dominant noise related
to magnetic fields is the interaction and down-conversion of test
mass eddy currents with time dependent magnetic fields, both originating
in the audio band. The current best estimate for LISA is that the
acceleration noise caused by eddy current damping and these magnetic
fields will be at $0.3\,\text{fm}/\text{s}^{2}\sqrt{\text{Hz}}$ above
a few mHz and then increases with $f^{-1}$. Possible mitigation strategies
could include $\mu$-metal and reductions in magnetic fields at audio
frequencies.

All these mitigation steps might allow to reduce the acceleration
noise in the high frequency region by an order of magnitude, as required
by ALIA, and will also increase the frequency band in which the acceleration
noise meets the LISA requirement of $3\,\text{fm}/\text{s}^{2}\sqrt{\text{Hz}}$.
For even lower frequencies, not surprisingly, temperature variations
are among the most crucial disturbances. Temperature fluctuations
and time dependent gradients push and pull on the test mass for example
through differential outgassing and differential radiation pressure
as well as the already discussed gravitational forces. Based on LPF
experience, LISA models assume a steep increase of the temperature
fluctuations towards lower frequencies $(f^{-3.5})$ starting with
$100\,\mu\text{K}/\sqrt{\text{Hz}}$ at $100\,\mu\text{Hz}$. However,
an active temperature sensing and control scheme could significantly
reduce the temperature fluctuations.

To summarize, now is the time to build on the extensive LPF experience
and use the lessons learned to improve acceleration noise beyond its
performance at higher frequencies and expand the performance to even
lower frequencies. The knowledge is fresh, the field is growing, and
starting the technology development now allows to develop a mature
mission concept for the next Decadal survey. But this requires a dedicated
technology development program which builds on the LPF experience
but should also evaluate alternative technologies and methods.

\subsection{Measurement Noise}

\begin{wrapfigure}{r}{0.6\textwidth}%
\noindent \centering{} \includegraphics[width=0.6\textwidth]{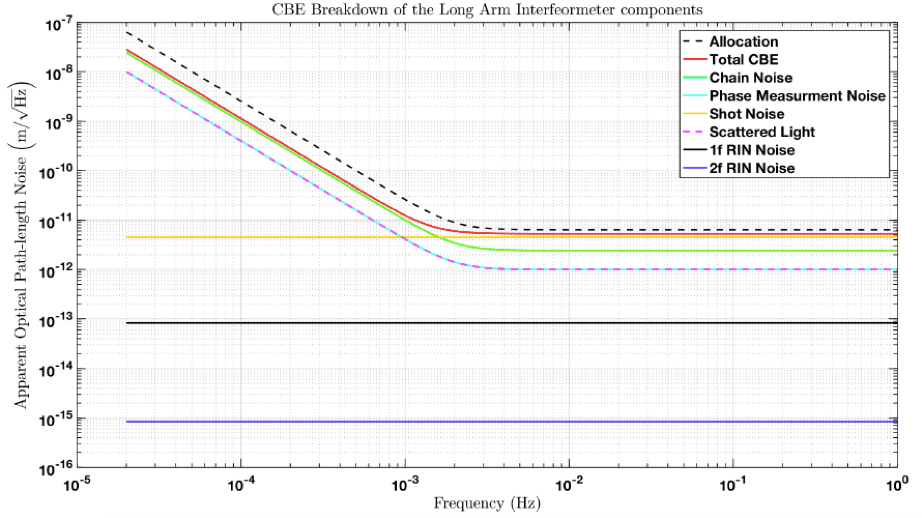}
\caption{The CBE of the most critical of LISA's long arm interferometer noise
sources \citep{LISAPerfModel}.}
\label{fig:densing_noise} \end{wrapfigure}

A second key technology is required to sense the minute changes in
distance between the two widely separated spacecraft. Laser interferometric
distance measurements are fundamentally limited by the intrinsic phase
noise which is inversely proportional to the amplitude of the metering
laser field. However, reaching this level is often very difficult
especially when the amplitude is large. Ground-based observatories
such as Advanced LIGO use optical cavities to amplify the response
and a Michelson interferometer to suppress the common parts of the
amplitude before they reach and saturate the detector. Both of these
techniques will be very difficult to implement between drag free spacecraft.
LISA is in that sweet spot where the received amplitude is at a level
which allows shot noise limited detection of $\sim10\,\text{MHz}$
beat signals with current technology. LISA still requires clock noise
transfer between the spacecraft and pilot tones to remove phase noise
added by the analog components. A sophisticated timing and ranging
system allows to suppress laser frequency noise using time delay interferometry.
All of these technologies have been demonstrated at the $\sim\text{pm}/\sqrt{\text{Hz}}$-level
in the LISA band sufficient to meet the LISA requirement of $10\,\text{pm}/\sqrt{\text{Hz}}$
equivalent single link displacement noise, twice as high as the allocated
shot noise limit. Figure\,\ref{fig:densing_noise} shows the breakdown
for the sensing noise in the long arm interferometer. It is dominated
by shot noise above 1\,mHz which will also limit the overall LISA
sensitivity at these frequencies. Technical noise sources are allowed
to increase with $f^{-2}$ below 1\,mHz where acceleration noise
is expected to dominate.

The Folkner mission uses hundred times longer arms which reduces the
received power by four orders of magnitude and increases the shot
noise limit by a factor 100, likely well below the galactic binary
background. If the overall performance of all technical noise sources
in the displacement measurement system continues to raise only with
$f^{-2}$ down to sub-$\mu$Hz, acceleration noise would be a factor
100 above these noise sources. This provides some needed margin to
accommodate expected faster increases in temperature fluctuations
towards the lower end of the $\mu$Hz spectrum which might otherwise
degrade the performance of the phase measurement chain. Note that
increases in the received laser power by either increasing the laser
power itself or the diameter of the telescopes is likely required
to enable for example inter-spacecraft communication or acquiring
lock. There are many other challenges related to this mission including
reaching the orbits, sunshades, and communication, however, the technical
challenges for the principle payload of such a low frequency mission
appear to be mostly related to acceleration noise.

The 200-fold improved shot limit for ALIA present significant challenges
for many subsystems within the interferometric measurement system.
LISA's photo detector noise and phase noise in the analog chain is
already state of the art for the $10-20\,\text{MHz}$ laser beat frequencies.
ALIA's orbit should allow for reduced Doppler shifts resulting in
lower laser beat frequencies if the intrinsic laser noise can be suppressed
below shot noise (1f-RIN-line) at these lower frequencies; the current
laser systems are not shot noise limited below $\sim8\,\text{MHz}$.
However, even if the relative intensity noise can actively be suppressed,
phase noise and timing jitter in the analog parts of the phase measurement
chain still need to be reduced by at least one order of magnitude.
This requires a dedicated effort to develop and study electrical,
electro-optical, and optical components starting from RF cables; the
temperature dependency of the electric susceptibility of the dielectric
inside the cable (often Teflon) is already a concern for phase noise
in LISA, to analog to digital converters and the timing of them. 

Beyond the individual components, the phase measurement system in
its entirety needs to be re-evaluated. Are strategies such as comparing
clock noise using the laser links still valid? What alternatives exist?
Are there designs which reduce the laser beat frequencies by a few
orders of magnitude to reduce demands on timing stability. What other
laser sources will be available in time for a new mission? Using a
lower laser wavelength decreases diffraction losses and increases
the phase shift. Furthermore, the amount of scattered light will increase
with the laser power while the sensitivity to scattered light increases
by a factor 200. This requires new strategies to reduce and/or cancel
stray light. Or generally speaking, to take full advantage of the
increased received laser power, LISA's typical $\sim\text{pm}$-requirements
for most technical noise sources in the measurement system turn into
$\sim5\,\text{fm}$-requirements above $\sim20\,\text{mHz}$. 

The feasibility of improvements at this magnitude need to be studied
and demonstrated in a dedicated technology development program before
a believable and mature bLISA mission concept can be developed and
presented to the next decadal survey.

\section{Organization, Partnerships, and Current Status}

We imagine that the technology development work proposed here could
be carried out within the structure of the ROSES program, in a manner
analogous to the Beyond Einstein Foundation Science (~2005-2007)
or LISA Preparatory Science (2018) calls. The directed calls could
be formulated by NASA Headquarters APD and PCOS staff with input from
the LISA project team and the concept study. The concept study would
be a panel of experts from the GW and astrophysics community similar
to the many panels created by APD over the last 2 decades. An initial
panel would identify promising concepts and the technology required
early in the decade. A second study would meet later in the decade
to evaluate the science return, the technology development progress
and the outstanding technology risk, in time to make a recommendation
to Astro2030, likely to commence in 2028.

\section{Schedule}

The goal of this proposal is to have a bLISA mission concept ready
for consideration by Astro2030. To accomplish this goal, the technology
challenges for candidate mission concepts have to be understood and
surmountable in order to select the concept. Hence, the technology
development work needs to proceed during the 2020s, accompanied by
mission concept studies to identify the requisite technologies and
support a selection. LISA technology development will continue into
the mid-2020s, when LISA goes into Phase C. The pace of technology
development for bLISA should be set to support a concept down-select
no later than 2028. That down-select will need to appraise the relative
challenges of the candidate missions and the technology development
path to flight for each. That could require only a low TRL, say 3-4,
i.e., proof of concept. After that time, technology development for
bLISA should focus on advancing toward TRLs of 5-6 in anticipation
of a Phase A start early in the 2030s. If LISA launches in 2034, as
currently planned, it would be finishing extended operations in the
mid-2040s. Optimally, a bLISA would launch in the latter half of that
decade.

\section{Cost}

This is a (very) small proposal for space related activities. At this
time, the technologies to be developed are not well-enough understood
to produce a detailed roadmap with schedule and budget profile. At
best, we can only offer ROM costs. Based on our experience with LISA
technology development, we estimate that the technology development
needed in the decade of the 2020s will be about \$20M. The cost of
a mission concept study will be about \$2M. The study might reasonably
span two episodes, an early study to identify the candidate concepts
and their requisite technologies, and then a later study ending by
2028 that evaluates the development progress, the development remaining
and the potential science return of the candidate mission concepts
in order to make the down-select. Then a proposal would be made to
Astro2030.

\pagebreak{}


\begin{thebibliography}{0}%
\makeatletter
\providecommand \@ifxundefined [1]{%
 \@ifx{#1\undefined}
}%
\providecommand \@ifnum [1]{%
 \ifnum #1\expandafter \@firstoftwo
 \else \expandafter \@secondoftwo
 \fi
}%
\providecommand \@ifx [1]{%
 \ifx #1\expandafter \@firstoftwo
 \else \expandafter \@secondoftwo
 \fi
}%
\providecommand \natexlab [1]{#1}%
\providecommand \enquote  [1]{``#1''}%
\providecommand \bibnamefont  [1]{#1}%
\providecommand \bibfnamefont [1]{#1}%
\providecommand \citenamefont [1]{#1}%
\providecommand \href@noop [0]{\@secondoftwo}%
\providecommand \href [0]{\begingroup \@sanitize@url \@href}%
\providecommand \@href[1]{\@@startlink{#1}\@@href}%
\providecommand \@@href[1]{\endgroup#1\@@endlink}%
\providecommand \@sanitize@url [0]{\catcode `\\12\catcode `\$12\catcode
  `\&12\catcode `\#12\catcode `\^12\catcode `\_12\catcode `\%12\relax}%
\providecommand \@@startlink[1]{}%
\providecommand \@@endlink[0]{}%
\providecommand \url  [0]{\begingroup\@sanitize@url \@url }%
\providecommand \@url [1]{\endgroup\@href {#1}{\urlprefix }}%
\providecommand \urlprefix  [0]{URL }%
\providecommand \Eprint [0]{\href }%
\providecommand \doibase [0]{http://dx.doi.org/}%
\providecommand \selectlanguage [0]{\@gobble}%
\providecommand \bibinfo  [0]{\@secondoftwo}%
\providecommand \bibfield  [0]{\@secondoftwo}%
\providecommand \translation [1]{[#1]}%
\providecommand \BibitemOpen [0]{}%
\providecommand \bibitemStop [0]{}%
\providecommand \bibitemNoStop [0]{.\EOS\space}%
\providecommand \EOS [0]{\spacefactor3000\relax}%
\providecommand \BibitemShut  [1]{\csname bibitem#1\endcsname}%
\let\auto@bib@innerbib\@empty
\end{thebibliography}%


\begin{thebibliography}{99}
\bibitem{LISAProposal2017}The LISA proposal is available at https://www.elisascience.org/files/publications/LISA\_L3\_20170120.pdf

\bibitem{aLIGO15}J. Aasi et al. (LIGO Scientific Collaboration),
Advanced LIGO, Classical and Quantum Gravity, Volume 32, Number 7
(2015)

\bibitem{GW150914}B.\LyXThinSpace P. Abbott et al. (LIGO Scientific
Collaboration and Virgo Collaboration), GW150914: First results from
the search for binary black hole coalescence with Advanced LIGO, Phys.
Rev. D 93, 122003 (2016)

\bibitem{GW170817}B.\LyXThinSpace P. Abbott et al. (LIGO Scientific
Collaboration and Virgo Collaboration), GW170817: Observation of Gravitational
Waves from a Binary Neutron Star Inspiral, Phys. Rev. Lett. 119, 161101 

\bibitem{GW_Plotter}Christopher J. Moore, Robert H. Cole, Christopher
P. L. Berry, Gravitational-wave sensitivity curves, Classical and
Quantum Gravity, Volume 32, Number 1

\bibitem{LPF18a}M. Armano, H. Audley, J. Baird, P. Binetruy, M. Born,
D. Bortoluzzi, E. Castelli, A. Cavalleri, A. Cesarini, and A. M. Cruise.
Beyond the Required LISA Free-Fall Performance: New LISA Pathfinder
Results down to 20 $\mu$Hz. Phys. Rev. Lett., 120(6):061101 ( 2018a). 

\bibitem{LPF18}G. Anderson, J. Anderson, M. Anderson, G. Aveni, D.
Bame, P. Barela, K. Blackman, A. Carmain, L. Chen, and M. Cherng.
Experimental results from the ST7 mission on LISA Pathfinder. Phys.
Rev. D, 98(10):102005 (2018). 

\bibitem{LPF18b}M. Armano, H. Audley, J. Baird, P. Binetruy, M. Born,
D. Bortoluzzi, E. Castelli, A. Cavalleri, A. Cesarini, and A. M. Cruise.
Precision charge control for isolated free-falling test masses: LISA
pathfinder results. Phys. Rev. D, 98(6):062001 (2018b). 

\bibitem{LPF19a}M. Armano, H. Audley, J. Baird, P. Binetruy, M. Born,
D. Bortoluzzi, E. Castelli, A. Cavalleri, A. Cesarini, and A. M. Cruise.
LISA Pathfinder platform stability and drag-free performance. Phys.
Rev. D, 99(8):082001 (2019a). 

\bibitem{LPF19b}M. Armano, H. Audley, J. Baird, P. Binetruy, M. Born,
D. Bortoluzzi, E. Castelli, A. Cavalleri, A. Cesarini, and A. M. Cruise.
Temperature stability in the sub- milliHertz band with LISA Pathfinder.
Accepted (2019b). 

\bibitem{LPF19c}M. Armano, H. Audley, J. Baird, P. Binetruy, M. Born,
D. Bortoluzzi, E. Castelli, A. Cavalleri, A. Cesarini, and A. M. Cruise.
Temperature stability in the sub- milliHertz band with LISA Pathfinder.
MNRAS, 486(3):3368–3379 (2019c). 

\bibitem{IPTA16}Verbiest, J. P. W., Lentati, L., Hobbs, G. et al.,
The International Pulsar Timing Array: First data release, MNRAS,
458, 1267 (2016)

\bibitem{Kagra}KAGRA collaboration, KAGRA: 2.5 generation interferometric
gravitational wave detector, Nature Astronomy 3, pages 35–40 (2019) 

\bibitem{EinsteinTelescope}Punturo, M. et al. The Einstein Telescope:
a third-generation gravitational wave observatory. Class. Quantum
Gravity 27, 194002 (2010).

\bibitem{3G}B. Abbott et al. for the LIGO Science Collaboration,
Exploring the sensitivity of next generation gravitational wave detectors,
Classical and Quantum Gravity, Volume 34, Number 4 (2017)

\bibitem{BBO2003}E. S. Phinney et al., Big Bang Observer, NASA Mission
Concept Study, 2003

\bibitem{CrowderCornish05}J. Crowder and N. Cornish, Phys Rev D 72,
083005 (2005)

\bibitem{ALIABender13}P. L. Bender, M. C. Begelman and J. R. Gair,
\char`\"{}Possible LISA follow-on mission scientific objectives\char`\"{},
Class. Quantum Grav. 30, 165017 (2013). 

\bibitem{NASASGO12}Gravitational-Wave Community Science Team, Gravitational-Wave
Core Team and Gravitational-Wave Science Task Force, Gravitational-Wave
Mission Concept Study Final Report, August 9, 2012. Available at https://www.lisa.nasa.gov/documentsReference.html. 

\bibitem{GADFLI}Sean McWilliams, Geostationary Antenna for Disturbance-Free
Laser Interferometry (GADFLI), arXiv:1111.3708 {[}astro-ph.IM{]}

\bibitem{gLISA}M. Tinto, Jose C.N. de Araujo, Coherent observations
of gravitational radiation with LISA and gLISA, Phys. Rev. D 94, 081101(R)
(2016)

\bibitem{LISAPerfModel}Martin Hewitson, Ewan Fitzsimons, Bill Weber,
LISA Performance Model and Error Budget, LISA-LCST-INST-TN-003
\end{thebibliography}
\end{document}